\documentclass[aps,prl,superscriptaddress,twocolumn]{revtex4-1}
\usepackage{graphicx}
\usepackage{amssymb}
\usepackage{amsmath}
\usepackage{psfrag}

\newcommand{\dert}{\partial_t}
\newcommand{\derx}{\partial_x}
\newcommand{\der}{\partial}

\begin{document}

\title{Quantum gray solitons in confining potentials}

\author{Dominic C.~Wadkin-Snaith}
\affiliation{School of Physics and Astronomy, University of Birmingham,
Edgbaston, Birmingham, B15 2TT, UK}
\author{Dimitri M.~Gangardt } 
\affiliation{School of Physics and Astronomy, University of Birmingham,
Edgbaston, Birmingham, B15 2TT, UK}

\ \\

\date{\today}

\begin{abstract} 
We define and study hole-like excitations (the Lieb II mode) 
in a weakly interacting Bose liquid subject to external confinement. 
These  excitations are obtained by  semiclassical quantization 
of gray solitons propagating on top of a Thomas-Fermi background.  
Radiation of phonons by an accelerated gray soliton leads 
to a finite life-time of these excitations. It is shown that, 
for a large number of trapped  atoms,  most 
of the Lieb II levels can be experimentally resolved.
\end{abstract}

\maketitle

Ultra cold atoms restricted to move in one dimension (1d) are now routinely
used for experimental realization and investigation of the physics of strongly
interacting many-body systems.  Due to advances in measurement techniques, the
dynamical response of one-dimensional quantum gases has recently gained a
central role in these studies
\cite{Kinoshita06,Hofferberth07,Koehl_PhysRevLett.103.150601}.

Dynamics is ultimately related to the nature of excitation spectrum.  The
latter is given by the Bethe Ansatz solution of Lieb-Liniger model
\cite{Lieb_Liniger_1963} for 1d ultra cold bosons with short-range interactions
\cite{Olshanii_PhysRevLett.81.938,Petrov2000Regimes}.  As shown by Lieb
\cite{Lieb_1963} the excitations can be decomposed into a superposition of
particle-like (Lieb I) and hole-like (Lieb II) excitations in one-to-one
correspondence with elementary excitations of one-dimensional
\emph{fermions}. For weak coupling,  the dispersion of Lieb~I
excitations can be obtained semiclassically by linearization of the
Gross-Pitaevskii Equation (GPE) resulting in phonon-like spectrum. 
Lieb~I excitations
determine the thermodynamic properties of the system and control the power-law
decay of correlation functions. 

Later, it was shown by Kulish, Manakov and Faddeev
\cite{KulishManakovFaddeev76} that the Lieb~II mode can be associated with
another semiclassical object -- the \emph{gray soliton}
\cite{Tsuzuki_1971}. The latter is a hole-like solution of GPE representing
localized density and supercurrent dip propagating on top a of constant
background. It is a mean field picture of many quantum Lieb~II excitations
forming a coherent wavepacket. This classical description holds as long as
quantum fluctuations remain small \cite{Dziarmaga2004} 
due to a large negative effective mass
resulting from the many particles expelled from the soliton core .  The same
large parameter is responsible for the exponentially suppressed probability to
create gray solitons using local perturbations
\cite{Khodas2008PhotosolitonicEffect}.

\begin{figure}[t]
 \centering
  \includegraphics[width=\columnwidth]{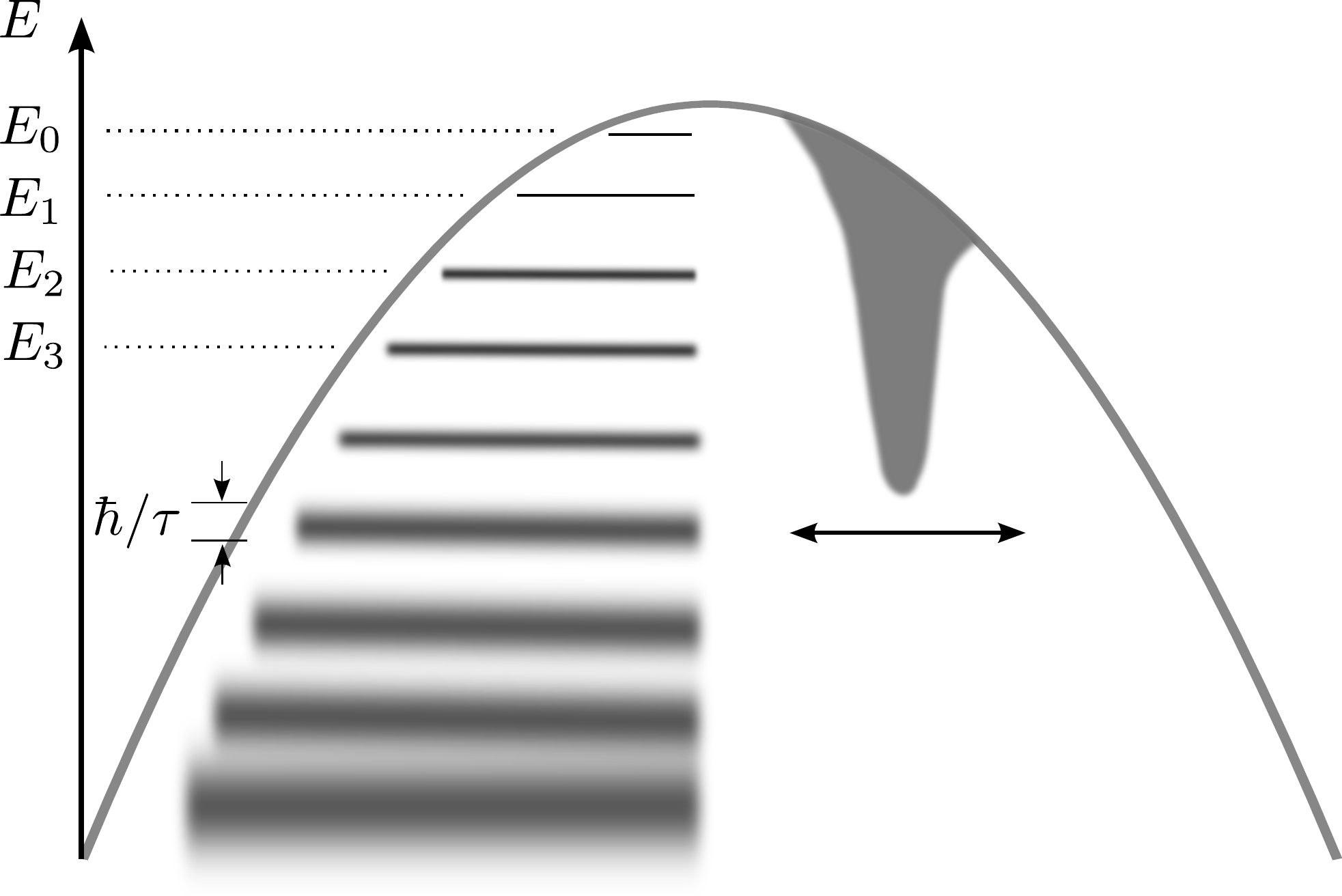}
 \caption{ Schematic picture of trapped Lieb II modes obtained by quantizing
   motion of gray solitons in the non-uniform background.  
   The non-uniform density
   profile leads to the finite life-time
   $\tau$ and  precludes resolution of low-lying levels.   }
 \label{fig:levels}
\end{figure}

Even a shallow axial trapping potential present in the experiments breaks the
integrability of the Lieb-Liniger model and the Lieb classification of the
excitations is apparently lost. While Lieb~I phonons can still be defined by
linearization of GPE around the appropriate density profile and quantizing the
resulting collective oscillations
\cite{menotti_stringari_2003,Gangardt2003Stability}, no analogous procedure
exists for the Lieb~II mode.  Classically, however, gray solitons do survive
the harmonic confinement if the latter is sufficiently smooth on the length
scale of the soliton \cite{Busch2000Motion,Konotop2004Landau}. They have
recently been observed in several experiments
\cite{Burger1999Dark,Anderson_2001_watching_dark_solitons,Becker08} in
agreement with theoretical studies
\cite{Muryshev2002Dynamics,Jackson2007Darksoliton}. The physics of 
underlying \emph{quantum} states has not been addressed yet. 

The aim of this Letter is to define and study the finite-size analogue of the
Lieb~II mode in a trapped system as corresponding \emph{quantum}
excitations. This is done by applying Bohr-Sommerfeld quantization rule to the
classical dynamics of gray soliton.  The obtained quasiparticle 
levels, shown schematically
in Fig.~\ref{fig:levels} have a finite life-time due to interactions of the
soliton with the background excitations.  The standard mechanism of soliton
decay due to the scattering of thermal phonons \cite{Muryshev2002Dynamics}
cannot be used here as it becomes inefficient for the exactly integrable GPE
\cite{Gangardt2010Quantum}. Even for non-integrable interactions the life-time
diverges as inverse fourth power of temperature $T$ \cite{castro96} and
solitons propagating in uniform background are absolutely stable at $T=0$ due
to conservation of energy and momentum.  The non-uniform density profile
relaxes momentum conservation and leads to soliton decay even at $T=0$.  We
calculate this fundamental limit on the life-time of the trapped quantum solitons.

Our main findings are as follows. For a system of $N_\mathrm{tot}$ interacting
bosons trapped in a potential $U(x)$ there are exactly $N_\mathrm{tot}$ quantum
states below the energy $E_\mathrm{ds}$ corresponding to the energy of
stationary classical dark soliton localized 
at the maximum of the trapped density profile.  
We associate these states, shown schematically in Fig. \ref{fig:levels},
with the trapped Lieb~II mode. In the case of harmonic confinement
$U(x)=m\omega^2x^2/2$ we use the Thomas-Fermi density profile to show that
$E_\mathrm{ds}=(\hbar \omega/\sqrt{2})N_\mathrm{tot} $ and the energies of the
trapped Lieb~II mode are given by the descending ladder
\begin{eqnarray}
  \label{eq:ladder}
  E_n = E_\mathrm{max}-\frac{\hbar \omega}{\sqrt{2}} 
  \left(n+\frac{1}{2}\right) \, ,
\end{eqnarray}
where $n=0,1,\dots,N_\mathrm{tot} -1$.
The energy of the highest \emph{quantum} state $n=0$ is  reduced 
by  zero-point oscillations $(\hbar\omega/\sqrt{2})/2$ from 
the classical value $E_\mathrm{ds}$. It is interesting to compare
this result with the case of extremely strong interactions, 
so-called Tonks-Girardeau limit \cite{Girardeau60}, 
where hard-core bosons can be mapped into
non-interacting fermions. In this case the Lieb~II mode  
is obtained by  creating holes in the filled Fermi
sea of $N_\mathrm{tot}$ particles occupying energies $E_n=E_\mathrm{F}-
\hbar \omega (n+1/2)$, below the Fermi energy
$E_\mathrm{F} =  \hbar \omega N_\mathrm{tot}$.

The trapped Lieb~II states are not eigenstates but rather  
quasiparticle resonances with finite life-time. 
The life-time for a state with energy $E$ is found to be
\begin{eqnarray}
 \label{eq:tau}
 \tau (E) =\frac{8\mu }{\hbar\omega^2}  F(x)\, ,   
\end{eqnarray}
where $\mu $ is the chemical potential of the condensate,
$x^3=E/E_\mathrm{ds}$, and the function $F(x)$ is defined in
Eq.~(\ref{eq:f}). For high energies, $E\sim E_\mathrm{ds}$, we find a
logarithmically large life-time $\omega\tau \simeq
(4\mu/\hbar\omega)\log\left[ 6E_\mathrm{ds}/(E_\mathrm{ds}-E)\right]$, while
for low energy states, $E \ll E_\mathrm{ds}$ we have $\omega\tau\simeq
(8\mu/3\hbar\omega)(E/E_\mathrm{ds})$. The last expression defines an energy
scale $E^* = (\hbar\omega/\mu) E_\mathrm{ds} $ below which $\omega\tau \le 1$
and Lieb~II quasiparticle states cannot be resolved. This condition coincides
with the classical picture of an overdamped  soliton decaying before
 completing one
period of oscillations in the trap. Another restriction, $E\gg E^{**} =
E_\mathrm{ds}/K^{3/2}=\mu/K^{1/2}$, arises from the requirement of the
validity of the semiclassical treatment
\cite{Khodas2008PhotosolitonicEffect}. Here $K=\pi E_\mathrm{ds}/\mu=(\pi \hbar
\omega /\sqrt{2}\mu) N_\mathrm{tot}$ is the Luttinger parameter in the center of
the trap.  For the experiment in Ref.~\cite{Becker08} one has $\hbar\omega/\mu \sim
10^{-2}$ and $K\sim 10^3$, therefore $E^* \simeq 0.01 E_\mathrm{ds} \gg
E^{**}$ and Eq.~(\ref{eq:ladder}) describes accurately the energies of the
most of the trapped Lieb~II states and most of these are well defined
quasiparticles, hence  our semiclassical quantization of the soliton 
is justified \emph{a posteriori}.

\emph{Solitons in uniform background.}---To 
derive the above results we consider the 
standard Lagrangian density for weakly interacting 1d bosons
$ \mathcal{L} = i\hbar\bar{\psi}\der_t\psi - 
  \mathcal{E}(\bar{\psi},\psi)$ with the energy density
  \begin{eqnarray}
    \label{eq:en_density}
    \mathcal{E} = \frac{\hbar^2}{2m} \left|\der_x\psi\right|^2 + 
    \frac{g}{2} \left(\left|\psi\right|^2-n\right)^2
  \end{eqnarray}
Here $\bar{\psi}(x,t),\psi(x,t)$ are bosonic fields, 
$m$ is the particle mass and $g>0$ characterizes repulsive interactions
between particles. In (\ref{eq:en_density}) we have subtracted the constant 
contribution  of the static background density $n$ (fixed by the 
chemical potential $\mu=gn$). Variation with
respect to the fields leads to GPE,
\begin{eqnarray}
i\hbar\der_t \psi = -\frac{\hbar^2}{2m} \der^2_x\psi +
g\left(\left|\psi\right|^2- n\right)\psi.
\end{eqnarray}
As shown in \cite{Tsuzuki_1971} 
it allows for a one-parameter family of gray solitons 
$\psi_\mathrm{s}(x-X(t))$
well localized  around the position $X$ 
moving with a constant velocity, $X=X_0+Vt$.
According to \cite{Tsuzuki_1971} the form of these
solutions is    
\begin{eqnarray}
 \label{eq:soliton}
\psi_\mathrm{s} (x) /\sqrt{n}= \cos\frac{\Phi}{2} -i\sin\frac{\Phi}{2}
\tanh\left( \frac{mgN }{2\hbar^2}x \right) .
\end{eqnarray}
Parameters $\Phi$ and $N$ are integral characteristics of the soliton
\cite{Pelinovsky1996Instabilityinduced} representing 
the total phase drop across the soliton and the number of particles expelled
from the soliton vicinity. The are both fixed by the velocity $V$ of the
soliton, 
\begin{eqnarray}
\label{eq:integralPhi}
 \Phi &=& 2\arctan\frac{\sqrt{c^2-V^2}}{V} \\
\label{eq:integralN}
 N &=&  \int\!dx \left(n-\left|\psi_\mathrm{s}(x)\right|^2\right)=
   \frac{2\hbar}{g}\sqrt{c^2-V^2}   .
\end{eqnarray}
Here $c=\sqrt{gn/m}$ is the sound velocity limiting the velocity 
of the soliton, $V^2<c^2$.

Following Ref.\cite{Konotop2004Landau} we wish to establish a description of a
soliton as an effective particle. To this end we substitute the solitonic
solution given by Eq.~(\ref{eq:soliton}) into Eq.~(\ref{eq:en_density}) and
integrate it over the length of the system to obtain the energy of the
soliton,
\begin{eqnarray}
\label{eq:energy_sol}
E_\mathrm{s} = \frac{4}{3}\frac{\hbar m}{g} \left(c^2-V^2\right)^{3/2}
=\frac{mg^2N^3}{6\hbar^2}\,. 
\end{eqnarray}
The canonical momentum of the soliton is 
\begin{eqnarray}
  \label{eq:momentum_sol}
P_\mathrm{s}= \hbar n\Phi -m N V  
\end{eqnarray}
The contribution $\hbar n \Phi$ comes from a small background supercurrent
\cite{Shevchenko1988} which carries no energy but must be introduced to
compensate for the phase drop $\Phi$.  The second contribution to the momentum
describes the deficit of $N$ particles moving with velocity $V$. Such an
expression for the momentum is completely general and independent of the details
of interactions between the particles
\cite{Schecter_Gangardt_Kamenev_arXiv1105.6136}. In contrast, the energy
$E_\mathrm{s}$ does depend on the form of interactions and the fact that it
only depends on the particle deficit $N$ is a direct consequence of the
quartic interaction of the GP energy density in Eq. (\ref{eq:en_density}).

For our purposes it is
convenient to use the canonical formalism where the momentum $P$, rather than
velocity $V$ serves as a control parameter. By formally inverting the relation
(\ref{eq:momentum_sol}) we obtain the velocity as a function of momentum,
\begin{eqnarray}
 \label{eq:velocity}
 \dot X = V(P,n)
\end{eqnarray}
Substituting $V(P,n)$ into Eqs.~(\ref{eq:integralPhi}),~(\ref{eq:integralN}) 
and Eq.~(\ref{eq:energy_sol}) yields 
$\Phi(P,n)$, $N(P,n)$ and $E_\mathrm{s} (P,n)$. 
The latter defines the dispersion of the soliton, or its effective Hamiltonian. 
Indeed, Eq.~(\ref{eq:velocity})  constitutes the Hamiltonian equation of motion 
as we have  $V = (\der E/\der V)/(\der P/\der V)=\der E/\der P$ 
following from Eqs.~(\ref{eq:energy_sol}) and
(\ref{eq:momentum_sol}). 
Another Hamiltonian equation of motion states the
conservation of momentum $\dot P =0$ expected  from the translational
invariance.   

\emph{Solitons in Thomas-Fermi density profile.}---Translational 
invariance is broken by the presence of external trapping
potential  $U(x)$. In this case the equation of motion for the momentum 
will be altered. It should be stressed, however, 
that the external potential does not act on the 
coordinate $X$ of the soliton directly, but rather on the particles
of the Bose gas forming the soliton. To describe this situation and derive
the equations of motion  we follow the method pioneered
in Ref.\cite{Konotop2004Landau}.  We assume the non-uniform 
background density is given by the Thomas-Fermi density profile
\begin{eqnarray}
\label{eq:TF}
g n (x) = \mu - U(x). 
\end{eqnarray}
For symmetric $U(x)$, a nonzero density requires $|x|<R$ 
where the  Thomas-Fermi radius $R$ is found from $U(R)=\mu$. 
The local sound velocity is  $c(x) = \sqrt{g n(x)/m}$.

For a sufficiently smooth potential and large $R$
the density $n(x)$ changes smoothly on the typical length scale 
of the soliton. In this case one can substitute $n\to n(X)$
into expressions for $\Phi(P,n)$, $N(P,n)$ and $E(P,n)$. The latter 
defines the effective soliton Hamiltonian in the non-uniform background, 
\begin{eqnarray}
 \label{eq:hamiltonianX}
H(P,X)=E(P,n(X))  \, ,
\end{eqnarray}
which depends on the external potential $U(X)$ only via the
Thomas-Fermi density profile (\ref{eq:TF}).

\emph{Classical and quantum dynamics of gray solitons.}---The 
Hamiltonian (\ref{eq:hamiltonianX}) generates the 
evolution for any dynamic observable $O(P,X)$
through the corresponding Poisson bracket
$  \dot O = \left\{ O, H \right\} =
 \der_X O\, \der_P H -
 \der_P O\, \der_X H . 
$
In particular,  the number of expelled particles $N$ 
is conserved during the dynamics as follows from Eq.~(\ref{eq:integralN}). 
The conserved combination with dimensions of energy
\begin{eqnarray}
\label{eq:h_eff}
mV^2-mc^2(X) =  m \dot X^2 + U(x) -\mu
\end{eqnarray}
can be used to solve for the dynamics of the soliton
by mapping onto an effective particle with mass
$2m$ moving in the potential $U(X)$  as was done in
Ref.~\cite{Konotop2004Landau}. Due to the positiveness of $c^2(X) - V^2$
the expression in Eq.~(\ref{eq:h_eff})  is always negative which 
restricts the motion to  the region where $U(X)<\mu$, 
\emph{i.e.} inside the atomic cloud $X<R$.

In what follows the  potential $U(x)$ is assumed to have
only one minimum situated at $X=0$. 
A classical point-like  trajectory $V=0$ and $X=0$
describes the stationary dark soliton with $\Phi=\pm \pi$ at the maximum 
of the density
profile corresponding to maximum energy~$E_\mathrm{ds}$.
Any trajectory with non-zero velocity $V$ has lower energy,
thus the classical energy is bounded between 0 and $E_\mathrm{ds}$.

\begin{figure}[t]
\centering
 \includegraphics[width=\columnwidth]{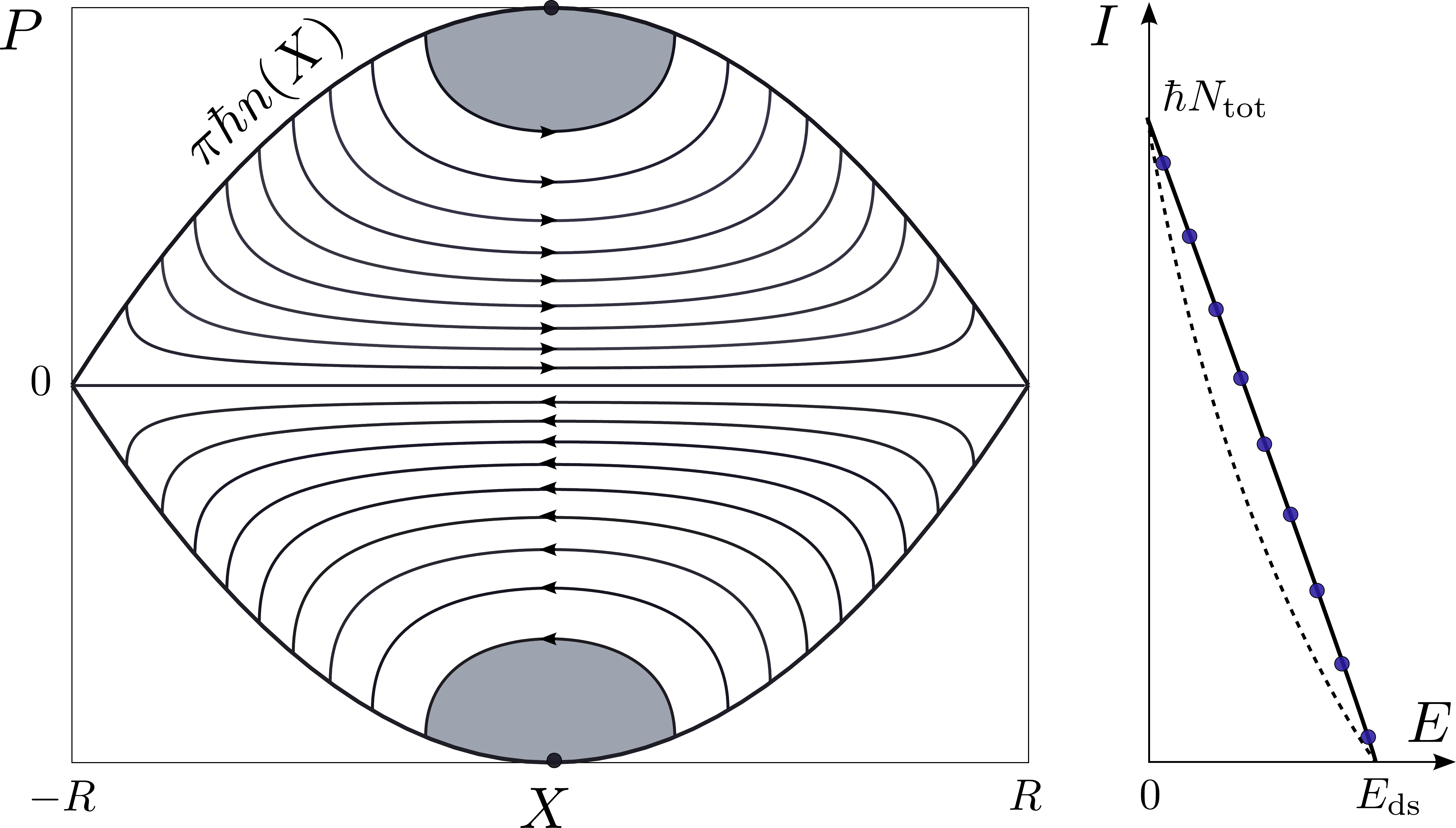}
 \caption{ The classical phase space trajectories of a trapped gray soliton
   for energies between $E=0$ (horizontal line for $P=0$) and
   $E=E_\mathrm{ds}$ (points at $X=0$ and $P=\pm \pi n_0$). The shaded area
   defines the value of action variable $I(E)$, Eq.~(\ref{eq:action_I}) which
   is shown as function of energy to the right. The dashed line corresponds to
   an arbitrary trap potential $U(x)$ while the straight solid line
   corresponds to an harmonic trap. Circles denote quantized values of
   $I(E)/\hbar = n+1/2$ given by the Bohr-Sommerfeld rule.  }
\label{fig:phase-space}
\end{figure}

The typical phase-space portrait is shown in Fig.~\ref{fig:phase-space}.  As
the velocity $V$ changes sign at the turning points, the phase $\Phi$ given by
Eq.~(\ref{eq:integralPhi}) undergoes a ``umklapp'' process and changes by
$2\pi$ which in turn results in a corresponding jump of the momentum in
Eq.~(\ref{eq:momentum_sol}). Calculating the oriented area enclosed by a given
orbit in the phase space defines the action variable,
\begin{eqnarray}
\label{eq:action_I}
I(E) = \frac{1}{2\pi} \oint P (E,X)\, dX\, .
\end{eqnarray}
as shown by shaded region in Fig.~\ref{fig:phase-space}.  Due to the hole-like
character of gray soliton excitations, the action is a \emph{decreasing}
function of the energy.  The maximum value of the action variable in 
Eq.~(\ref{eq:action_I}) corresponds to the singular classical orbit which
encloses the entire available phase space and on its two branches $\Phi=0,2\pi
$, $V=\pm c$ and $N=0$.  For this orbit the action variable takes the form
\begin{eqnarray}
  \label{eq:imax}
I_\mathrm{max} = \int dX \hbar n(X) = \hbar N_\mathrm{tot}\, .  
\end{eqnarray}
The minimum
value, $I=0$ is realized for the point-like trajectory with $E=E_\mathrm{ds}$.
For intermediate energies the integral in Eq.~(\ref{eq:action_I}) is a smooth
function decreasing from $I(0)=\hbar N_\mathrm{tot}$ to $I(E_\mathrm{ds})=0$.
The classical frequency of oscillations is given by $dE/dI=\Omega(E)$ which is
negative as expected for the hole-like excitations.

Applying the standard Bohr-Sommerfeld semiclassical quantization rule, the
energy levels are found from the condition $I(E_n)/ \hbar = n+1/2$. Therefore
there are exactly $N_\mathrm{tot}$ quantum states, \emph{for any trapping
  potential} $U(x)$.  In the case of an harmonic confining potential
$U\left(X\right) = m\omega^{2}X^{2}/2$, the effective energy in
Eq.~(\ref{eq:h_eff}) describes harmonic motion with energy-independent
oscillation frequency
$\Omega=-\omega/\sqrt{2}$~\cite{Busch2000Motion,Konotop2004Landau}, hence
$E_\mathrm{ds} = (\hbar\omega/\sqrt{2})N_\mathrm{tot}$ and we obtain
equidistant spectrum given by Eq.~(\ref{eq:ladder}).

\emph{Interactions with phonons.}---The picture presented above of solitons as
Landau quasiparticles must be complemented with their interactions with the
phonons. Indeed, a soliton propagating in a smooth non-uniform density profile
is analogous to an electron moving in static electric field created by some
external sources. It is well known that its deceleration leads to radiation of
electromagnetic waves removing momentum and energy from the electron, known as
\emph{Bremsstrahlung} \cite{Landau_Lifshitz_2_Classical_Theory_of_Fields}.  In
the case of the gray soliton the emitted phonons lead to an eventual loss of
energy and momentum from the soliton resulting in its acceleration due to
its negative effective mass.  To describe such a process we consider a combined
system of the soliton and phonons
\begin{eqnarray}
\label{eq:full_lagrangian}
L_\mathrm{eff}=L_\mathrm{s}+L_\mathrm{s-ph}+L_\mathrm{ph}
\end{eqnarray}
where $L_\mathrm{s}=P\dot X - H(P,X)$ and  
the last term describes the low-energy phonons in the quadratic harmonic
approximation \cite{PopovBookFunctional,HaldanePRL81},
\begin{eqnarray}
\label{eq:l_ph}
L_\mathrm{ph} = \int\!dx\,\left[ -\hbar\rho\dert\varphi -\frac{mc^2}{2n} \rho^2 
-\frac{\hbar^2n}{2m}\left(\derx\varphi\right)^2 \right]
\end{eqnarray}
Here we use 
density-phase representation for slow bosonic fields 
$\psi(x,t) =\sqrt{n+\rho(x,t)} \exp(i\varphi(x,t))$  
\footnote{We model the phononic
bath by homogeneous system, therefore the parameters in Eq.~(\ref{eq:l_ph}) 
should be calculated in the center of the trap, \emph{i.e.} for $x=0$. As long
as there is no reflection of phonons from the edges of the trap our results 
are unaffected by this choice.}.
The coupling $L_\mathrm{s-ph}$ between soliton and low-energy phononic modes
is given by the universal form,
\begin{eqnarray}
\label{eq:unicoupling}
 L_\mathrm{s-ph}=  -\hbar
\dot{\Phi}\,\vartheta(X,t)/\pi-\hbar
\dot N\,\varphi(X,t), 
\end{eqnarray}
where we have introduced displacement field 
$\vartheta(x,t) = \pi \int^x \rho(y,t) dy$. 
The universal coupling (\ref{eq:unicoupling}) was derived in
Ref. \cite{Schecter_Gangardt_Kamenev_arXiv1105.6136} directly from the 
principle of 
translational and gauge invariance and was shown to remain valid away 
from weak coupling limit. It shows that it 
is the soliton's total phase and the number of
particles ejected from the soliton core (or, rather their temporal change) 
that couple to the phonon fields.  
The coupling (\ref{eq:unicoupling}) is also local, involving 
phonon fields at the position of the soliton.


\emph{Dissipative dynamics of gray solitons.}--- The phonon action in
Eq~(\ref{eq:l_ph}) is quadratic, thus we are able to directly integrate out
the phononic degrees of freedom using the Keldysh formalism
\cite{Schecter_Gangardt_Kamenev_arXiv1105.6136} to obtain an effective action
for the soliton.  We obtain the following equations of motion modified by the
phonons
\begin{eqnarray}
\label{eq:dotX}
\dot X &=& V + \frac{\hbar\kappa}{2} 
\dot{\Phi}\,\der_P \Phi+
\frac{\hbar}{2\kappa}\dot N\,\der_P N  \\
\label{eq:dotP}
\dot P &=&  - \der_X H - 
\frac{\hbar\kappa }{2}\dot{\Phi}\,\der_X \Phi
-\frac{\hbar}{2\kappa}\dot{N}\,\der_X N \nonumber  \\
&&-  \frac{\hbar c}{c^{2} - V^{2}}\left(\frac{\kappa V}{2c}\dot{\Phi}^{2} + \dot{\Phi}\dot{N} + \frac{V}{2\kappa c}\dot{N}^{2}\right).
\end{eqnarray}
Here $\kappa=\hbar n/mc$ is a large number related to the Luttinger parameter
$K=\pi\kappa$ of the phononic Lagrangian (\ref{eq:l_ph}) and  
$V=\der_P H$ is the velocity of the soliton in the absence of 
phonons. These equations hold in the adiabatic limit, assuming
slow deviation of soliton parameters. Similar
expressions were obtained in Ref.\cite{Pelinovsky1996Instabilityinduced}.

As is well know from the theory of radiation in QED (see \emph{e.g.}
Refs.\cite{Landau_Lifshitz_2_Classical_Theory_of_Fields,
  Ginzburg_Applications_of_Electrodynamics}) the dissipative
Eqs.~(\ref{eq:dotX}) and (\ref{eq:dotP}) are plagued with runaway solutions.
The only consistent way to treat these equations is to solve for the soliton
trajectory $P(t)$, $X(t)$, $N(t)$, $\Phi(t)$ in the absence of phonons and
calculate the non-adiabatic corrections perturbatively.  As the number of
particles $N$ ejected from the soliton is conserved the r.h.s. of
Eqs.~(\ref{eq:dotX}) and (\ref{eq:dotP}) simplifies considerably.  Turning to
the energy dissipation rate, we have
\begin{equation}
\label{eq:dot_H}
\dot{H} = \dot P \der_P H +\dot X \der_X H=
 -\frac{\hbar\kappa}{2} \frac{c^{2}}{c^{2}-V^{2}}\dot{\Phi}^{2}. 
\end{equation}
Using Eqs.~(\ref{eq:integralPhi}),(\ref{eq:integralN}) 
one can show that  $\dot \Phi=-gN\dot V /c^2$. Averaging Eq.~(\ref{eq:dot_H})
over one period using harmonic motion 
$V/c =\sqrt{1-(N/2\kappa)^2}\cos(\omega t/\sqrt{2})$,
we find the rate of slow change of energy of the soliton  
\begin{eqnarray}
\dot E &=&-\frac{1}{T}\int^{T}_{0} \dot{H} dt = -\frac{2\hbar\kappa}{c^2}
\frac{1}{T} \int^{T}_{0} \dot V^2 dt\nonumber \\ 
&=& \frac{\hbar\kappa \omega^2 }{2}\left[1-\left(\frac{N}{2\kappa}\right)^2\right] 
=\frac{mg^2N^2 }{2\hbar^2}\dot N    \, .   
\end{eqnarray}
This allows us  to calculate the life-time  of the soliton,
\begin{eqnarray}
 \label{eq:tau_N}
 \tau \!\!&=&\!\! \int_0^N\!\! \frac{dN}{\dot N}\! =\! 
\frac{ mg^2}{\hbar^3 \kappa \omega^2}\!\int_0 ^N
\!\!\!\!\frac{N^2\, dN }{1\!\!-\!\!(N/2\kappa)^2}
\!\!= \!\!\frac{8\mu}{\hbar\omega^2} F\!\left(\frac{N}{2\kappa}\right),
\end{eqnarray}
where we have defined
\begin{eqnarray}
 \label{eq:f}
 F(x) =\int_0^{x} \frac{y^2\, dy}{1-y^2} 
 = \frac{1}{2}\log\frac{1+x}{1-x}-x\, .
\end{eqnarray}
For small $x$ it behaves like $F(x)\simeq x^3/3$ while near $x=1$ it diverges
logarithmically $F(x) \simeq \log\sqrt{2/(1-x)}$. Using the fact that 
$(N/2\kappa)^3 = E/E_\mathrm{ds}$  we get
the results stated in and after Eq.~(\ref{eq:tau}).

\emph{Conclusions.}---Hole-like Lieb~II excitations can be
generalized for trapped systems by quantizing dynamics of gray
soliton in non-uniform background. Despite their finite life-time, the obtained
excitations are well defined for current experimental conditions. This
can be verified by studying spectrum of phonon absorption, \emph{e.g} using  
parametric driving proposed recently in Ref.\cite{Proukakis2004}. To reveal
the quantum nature of the soliton its Heisenberg uncertainty 
$\Delta X = \sqrt{\hbar/8 m\omega  N}$ in the position 
has to be larger than its width, $\xi=2\hbar^2/mgN$ \footnote{The thermal
broadening can be neglected as for $T\ll \mu$ it is much smaller than the 
witdth of the soliton.}. Their  ratio $\Delta X/\xi =(\sqrt{\pi}/{4})
\sqrt{\mu/\hbar\omega K}\sim
 \left(mg^2 /\hbar^3\omega\right)^{1/3}
N^{1/6}_\mathrm{tot}$ is already $0.14$ for experiment in Ref.\cite{Becker08},  
and can be increased,  \emph{e.g} by reducing $\omega$ and  
making interactions $g$ stronger by a
tighter transverse confinement \cite{Olshanii_PhysRevLett.81.938}.

\emph{Acknowledments.}---
We are grateful to V.~Cheianov, A.~Kamenev, M.~Schecter, I.V.~Lerner, 
K.~Bongs and
L.I.~Glazman for fruitful discussions and J.M.F.~Gunn for critical comments.   
D.C.W.-S. acknowledges an EPSRC  studentship.
D.M.G. acknowledges EPSRC Advanced Fellowship EP/D072514/1.

\bibliography{../../References/references}

\end{document}